\documentclass[aoas,nameyear,MSNbibl,dvips]{arximspdf}
\usepackage{dcolumn}
\usepackage{graphicx}

\doi{10.1214/10-AOAS444}
\volume{5}
\issue{2B}
\pubyear{2011}
\firstpage{1678}
\lastpage{1698}

\makeatletter
\newcolumntype{d}[1]{D{.}{.}{#1}}

\newcommand{\bt}[1]{\bolds#1}
\def\bbeta{\bolds\beta}

\def\bptnote#1{}

\makeatother

\begin{document}
\begin{frontmatter}

\title{Bayesian Synthesis: Combining subjective analyses, with an application to ozone data\thanksref{T1}}
\runtitle{Bayesian synthesis}
\thankstext{T1}{This material is based upon work supported by the NSF through
     Awards  SES-0214574, SES-0437251, DMS-06-05041, and DMS-06-05052
     and by the NSA through Award MSPF-04G-109.}

\begin{aug}
\author[A]{\fnms{Qingzhao} \snm{Yu}\ead[label=e1]{qyu@lsuhsc.edu}},
\author[B]{\fnms{Steven N.} \snm{MacEachern}\ead[label=e2]{snm@stat.osu.edu}}
\and
\author[B]{\fnms{Mario} \snm{Peruggia}\corref{}\ead[label=e3]{peruggia@stat.osu.edu}}
\runauthor{Q. Yu, S. N. MacEachern and M. Peruggia}
\affiliation{Louisiana State University
Health Sciences Center, Ohio State University and Ohio State University}
\address[A]{Q. Yu\\
School of Public Health\\
Louisiana State University Health\\\quad Sciences Center\\
   New Orleans, Louisiana 70122\\
   USA\\
  \printead{e1}} 
\address[B]{S. MacEachern\\
   M. Peruggia\\
   Department of Statistics\\
   Ohio State University\\
   Columbus, Ohio 43210\\
   USA\\
   \printead{e2}\\
   \phantom{E-mail: }\printead*{e3}}
\end{aug}

\received{\smonth{8} \syear{2009}}
\revised{\smonth{11} \syear{2010}}

%
\begin{abstract}
Bayesian model averaging enables one to combine the disparate
predictions of a number of models in a coherent fashion, leading to
superior predictive performance. The improvement in performance
arises from \mbox{averaging} models that make different predictions. In this
work, we tap into perhaps the biggest driver of different
predictions---different analysts---in order to gain the full benefits
of model averaging. In a standard implementation of our method,
several data analysts work independently on portions of a data set,
eliciting separate models which are eventually updated and combined
through a~specific weighting method. We call this modeling procedure
Bayesian Synthesis. The methodology helps to alleviate concerns about
the sizable gap between the foundational underpinnings of the Bayesian
paradigm and the practice of Bayesian statistics. In experimental
work we show that human \mbox{modeling} has predictive performance superior
to that of many automatic modeling techniques, including AIC, BIC,
Smoothing Splines, CART, Bagged CART, Bayes CART, BMA and LARS, and
only slightly inferior to that of BART. We also show that Bayesian
Synthesis further improves predictive performance. Additionally, we
examine the predictive performance of a simple average across
analysts, which we dub Convex Synthesis, and find that it also
produces an improvement. Compared to competing modeling methods
(including single human analysis), the data-splitting approach has
these additional benefits: (1) it exhibits superior predictive
performance for real data sets; (2) it makes more efficient use of
human knowledge; (3) it avoids multiple uses of the data in the
Bayesian framework: and (4) it provides better calibrated assessment of
predictive accuracy.
\end{abstract}

%

\end{frontmatter}

\section{Introduction}\label{sec1}

A coarse but conceptually useful taxonomy of modeling strategies
distinguishes between two broad categories: automatic strategies and
strategies which require human intervention. Automatic strategies
typically rely on generic methods for model selection, perhaps
allowing data-based choice of a couple of tuning parameters. They are
appealing because, once the data are input, inferences are produced
without requiring any further human interaction. By contrast, human
modeling emphasizes exploratory data\vadjust{\goodbreak} analysis and the accompanying
notions of model development and refinement. The debate on the
relative merits of these two approaches is vigorous and ongoing [see,
e.g., Breiman
(\citeyear{Breiman2001}) or Hand
(\citeyear{Hand2006}), and the ensuing comments
and rejoinders].

In our experience, much of data analysis is heavily based on
subjective decisions which do not lend themselves to routine
formulations. These range from what variables to include in an
analysis to what forms the variables should take, to insight about the
parametric form of the response variable, to whether individual cases
should be included in the analysis or trimmed as outliers. Many
common instances of human interventions in the modeling cannot be
easily carried out by automatic procedures.

Throughout, an adequate analysis must take into account what the
variables are, whether they are well measured or of lesser quality,
whether individual influential cases drive the results, what the
scientific background of the problem is, etc. [Weisberg (\citeyear{Weisberg1985})]. All
of these elements are essential, both when modeling the data formally
and when drawing conclusions from the analysis. Also, in certain
cases, we might specify some aspects of a model and impose specific
constraints based on scientific knowledge that a general purpose model
selection method may fail to recognize.

Because of these reasons, we strongly adhere to the belief that a good
data analysis based on \textit{human intervention} will often be far
superior to a~routinely implemented analysis. In this article we
present a modeling and weighting strategy, called \textit{Bayesian
Synthesis}, for combining analyses from several human modelers
within the Bayesian framework. Bayesian Synthesis, formalized in
Section~\ref{framework}, relies on a number of different analysts each
contributing a Bayesian model to a pool of models. Each model in the
pool is given a weight, thus creating a ``hyper-model.'' The
techniques of model averaging [e.g., Raftery, Madigan and Hoeting (\citeyear{Raftery1997})] are used to
synthesize the different analysts' beliefs. Formal rules ensure that
the analysts will contribute models that can be synthesized. Bayesian
Synthesis retains the benefits of subjective modeling while
substantially enhancing the inferential and predictive strengths of
each individual analysis, producing combined inferences that vastly
outperform inferences based on automatic methods.

The methodology we propose can be viewed as a means of constructing a
useful space of models over which to perform a Bayesian analysis. In
this regard, it is strongly connected to the literature on model
selection [e.g., George
and McCulloch (\citeyear{George1993}), who describe a method of
screening models for further development] and on accounting for model
uncertainty [see Draper
(\citeyear{Draper1995}) and the following discussion for an
extensive treatment]. In contrast to earlier work, our approach
emphasizes the role of subjective modeling and the need for multiple
analysts.

In this article we report on the experimental development of the new
methodology. Specifically, we have constructed a careful experiment
(with appropriate randomization and blinding) that allows us to
contrast subjective modeling, and \mbox{subjective} modeling combined with
Bayesian Synthesis and Convex Synthesis, to automated modeling
methods. The results demonstrate the success of our new methods: With
only the exception of BART [Chipman,\vadjust{\goodbreak} George and McCulloch (\citeyear{Chipman2010})], subjective, human
modeling had predictive performance superior to that of a variety of
automatic methods, including AIC [Akaike
(\citeyear{Akaike1974})], BIC [Schwarz
(\citeyear{Schwarz1978})],
Smoothing Splines [Craven
and Wahba (\citeyear{Craven1979}); Gu
(\citeyear{Gu2002})], CART [Breiman et al. (\citeyear{Breiman1984})], a~bagged version of CART [Breiman
(\citeyear{Breiman1996})], LASSO
[Tibshirani
(\citeyear{Tibshirani1996})], Forward Stagewise
[Hastie, Tibshirani and Friedman (\citeyear{Hastie2001})], LARS
[Efron
et al. (\citeyear{Efron2004})], Bayesian Model Averaging [Raftery, Madigan and Hoeting (\citeyear{Raftery1997})]
and Bayesian CART [Chipman, George and McCulloch (\citeyear{Chipman1998})]. The gains relative to these
methods were large. The comparisons with BART give a~slight advantage
to BART, but not uniformly so. Bayesian Synthesis and Convex
Synthesis provide an additional, modest improvement over subjective
modeling. In addition, and much more importantly, it leads to a~more
realistic assessment of predictive accuracy, curbing the over-optimism
of each individual analyst.

In Section~\ref{framework} we introduce a Bayesian framework for data
splitting and formally describe Bayesian Synthesis and Convex
Synthesis. In Section~\ref{application} we present the experiment and
a careful discussion of the results.
In Section~\ref{discussion} we discuss related work and suggest
directions for future research.

\section{A Bayesian framework for data-splitting}\label{framework}
Our primary focus is on Bayesian modeling, where a team of analysts
builds models for a data set. The paradigm we envision is this.
First, the data are split into several portions. Each analyst
receives one portion of the data. Second, each analyst builds a
Bayesian model for their portion of the data, reporting a
``Bayesian summary'' of their posterior distribution. Third,
the Bayesian summaries are updated on portions of the data not
used to build them, and they are combined to yield a single, overall
posterior model.

Two features are essential for this procedure to work well. First,
each analyst must produce a Bayesian summary that is amenable to
updating with further data. Second, the various Bayesian summaries
must be amenable to synthesis. Throughout, we must exercise care
so that the data are not split into too many parts.
We will assume that there are $k$ analysts.

\subsection{Splitting the data}
The data to be used for model development and synthesis are split into
$k$ portions. Once split, the portions of the data are assigned to
the $k$ analysts at random. This produces an exchangeable partition
and assignment of data to analysts. Theoretical results presented in
Yu
(\citeyear{Yu2006}) suggest that (where data splitting is appropriate) the
portions of the data should all contain approximately the same amount
of information about the data-generating process. Following this
theory, we seek to produce a set of splits that give conditionally
i.i.d. data to the analysts. The following cases describe two of the
splitting procedures that we have implemented.

The first case is that of a designed experiment where a
structural balance is
forced upon the data. For example, the two-sample,\vadjust{\goodbreak} completely randomized
design is often implemented in a balanced fashion, so that the same
number of experimental units are assigned to each of the two treatment
conditions.
Additionally,
covariates are recorded on the experimental units.
For this type of experiment, we split at
random, with the restriction that each analyst receive the same number
of observations on each treatment. The additional covariates need not
be balanced and need not be used by the analysts in constructing a model
for the data.

The second case, matching the ozone example of
Section~\ref{application}, is one where there is a collection of
experimental units, with a variety of information on each unit. In
this case, we split the data at random, with each analyst receiving
the same number of observations.

These methods of splitting the data have the advantage of not
depending on the analysts' eventual models---an essential part of our
paradigm. The methods are extremely easy to implement and do not
require the help of an expert to split the data. The drawback to
these methods is that the portions of the data will typically not
convey the same amount of information to the different analysts.
While ``optimal'' splits might well differ, we would need to know the
details of the analysts' models to formalize the notions of
information in the splits and of optimality. For large samples, the
splits of the data will contain approximately the same amount of
information.

\subsection{Building and updating the model}
In order to carry out the analysis, each analyst is provided with a
set of ground rules for model building. The rules include, most
importantly, the goals of the modeling task. Second, the analyst must
know what kind of Bayesian summary to produce. Since the Bayesian
synthesis of the analysts' summaries will be accomplished through
Bayes factors, and since Bayes factors depend on the marginal
likelihood of the data, the analyst must be informed of the quantity
for which the likelihood will be calculated. Third, the analyst must
know what conventions will be followed for computation of the
likelihood. These conventions must guarantee that the analysts'
models will be mutually absolutely continuous over the range of values
that the data can assume.

Consider the prototypical experiments for which data splitting is
described. In the first case, of a balanced two-sample experiment
with case-specific covariates,
interest may focus on the difference between treatment means.
Implicitly, the analysts have been informed that the treatment means
exist. The Bayesian summary for an analyst represents the analyst's
posterior, given the portion of the data used for the analysis. The
likelihood of responses to the two treatments will be computed; the
mechanism assigning units to the treatments will not be part of the
likelihood. The convention for the likelihood is that it be a density
absolutely continuous with respect to  Lebesgue measure with support on
the real line. An alternate convention might be that the likelihood
be discrete, rounded to a single decimal place, on the
nonnegative half-line.

An instance of the second case is described in some detail in the
upcoming example, and so we leave off discussion for the moment. In
any event, each analyst is left with the choice of constructing a
model from the assigned portion of the data.\vadjust{\goodbreak} The analysts may use any
method whatsoever to build their model, ranging from automated
methods, to subjectively elicited priors, to construction and
refinement of models through diagnostics. The essence of the paradigm
is to encourage the analysts to build creative models that can be
combined across analysts.

\subsection{The Bayesian summary}
The Bayesian summary can take on a wide variety of forms, depending
on the analyst's modeling choices. Whatever
the form, the summary must be amenable to updating and allow one to
compute the marginal likelihood for the portions of the data not used
to construct the model.

Several forms of summary work well in practice. Choice of a
posterior distribution conjugate to the analyst's
chosen likelihood for the future data leads to a direct computation
of the marginal likelihood. Choice of a mixture of such distributions
leads to a mixture of conjugate posteriors, and hence to quick computation
of the marginal likelihood.
For models that move beyond conjugacy, the posterior distribution can
be represented in a discrete fashion, for example, by the output of a
Monte Carlo simulation. Along with the representation, the summary must
include a means of updating the summary, for example, code to compute the
marginal likelihoods and to produce summaries that enable one to address
the inferential goals of the analysis.

\subsection{Synthesizing the analyses}
When each analyst has produced a model, we can combine them to yield
an overall model. Under Bayesian Synthesis, we combine the models by
computing pairwise Bayes factors for portions of the data and then
reconciling them through the calculation of the geometric mean of
pairwise Bayes factors for each analyst. These geometric means
determine the weight that each analyst receives in predictions. A
formal justification for this choice of weighting is provided in
Sections~\ref{mw_justify} and~\ref{mw_unique}.

Let $Y_1, \ldots , Y_k$ denote the $k$ splits of the data; let $f_1,
\ldots , f_k$ denote the likelihoods for the $k$ models with possibly
differing parameters $\bt{\theta}_1, \ldots , \bt{\theta}_k$. The
pairwise Bayes factor is computed on the greatest set of data not used
in constructing the two models, after the two models have been updated
to include the same data. Thus, the Bayes factor comparing analysts
$1$ and $2$~is
\[
B_{12} = \frac
{ \int f_1(Y_3, Y_4, \ldots , Y_k | \bt{\theta}_1)
\pi(\bt{\theta}_1 | Y_1, Y_2)\,d\bt{\theta}_1}
{ \int f_2(Y_3, Y_4, \ldots , Y_k | \bt{\theta}_2)
\pi(\bt{\theta}_2 | Y_1, Y_2)\,d\bt{\theta}_2}
= \frac{m_{1(2)}}{m_{2(1)}}.
\]
Note that the distribution on $\bt{\theta}_1$ used in the above
calculation is the posterior, given both $Y_1$ and $Y_2$. Similarly,
the distribution on $\bt{\theta}_2$ is the posterior given both $Y_1$
and $Y_2$.

If the Bayesian summaries yield models that are each well represented
by a set of $N$ draws from the appropriate posterior distribution, the
Bayes factor can be estimated as
\[
\widehat{B}_{12} = \frac
{\sum_{j=1}^N N^{-1} f_1(Y_3, Y_4, \ldots , Y_k | \bt{\theta}_1^{(j)})}
{\sum_{j=1}^N N^{-1} f_2(Y_3, Y_4, \ldots , Y_k | \bt{\theta}_2^{(j)})}.
\]
Weighted distributions, such as those produced by importance sampling,
can be used to obtain the Bayes factor. For more complex models,
sophisticated methods of estimating the marginal likelihoods produce
these Bayes factors. See Chen
et al. (\citeyear{Chen2000}) for a recent book that
describes methods for estimating Bayes factors/marginal likelihoods.

Next, for each $i$, we compute the geometric mean of the estimated
Bayes factors to obtain
%
\begin{equation}\label{eq:weight_synthesis}
b_i =  \Biggl[\prod_{l=1}^k
\widehat{B}_{il} \Biggr]^{1/k},
\end{equation}
where $\widehat{B}_{ii} \equiv1$. These $b_i$ are then used as
weights to yield the synthesized posterior: $f(\bt{\theta}| Y) =
\sum_{i=1}^k b_i f(\bt{\theta}_i | Y_1, \ldots , Y_k)
 /\sum_{j=1}^k b_j  $. In this expression, $\bt{\theta}$~%
runs over the parameter spaces for all of the analysts' models.

\subsection{Model weights: A formal justification} \label{mw_justify}
Forecasts are naturally combined through the marginal likelihood. In
the context of model averaging performed by a single analyst, this
follows from Bayes theorem: assuming that equal prior weight is
assigned to each submodel under consideration, the posterior weight
for a submodel is then proportional to the Bayes factor for that
submodel against an arbitrary reference submodel. Thus, the ratio of
the weights assigned to two submodels equals the Bayes factor for one
against the other, and the Bayes factor expresses the impact that the
data have on the relative weights assigned to two submodels.

The approach that we have taken extends the result for a single
analyst to more than one analyst. When there are two analysts, each
plays the role of a submodel, and from the definition of
equation~(\ref{eq:weight_synthesis}), we have
\[
\frac{b_1}{b_2} =
\frac{  (\widehat{B}_{11}\widehat{B}_{12} )^{1/2} }
{  (\widehat{B}_{21}\widehat{B}_{22} )^{1/2} } =
 (\widehat{B}_{12} )^{1/2} (\widehat{B}_{12} )^{1/2}
= \widehat{B}_{12}.
\]
The formula for the $b_i$ given in
equation (\ref{eq:weight_synthesis}) does appear to be unusual, but it
produces the answer we had hoped for: the ratio of the weights equals
the Bayes factor. This formula for two analysts is used in the
analysis of the ozone data presented in Section~\ref{application}.

When there are more than two analysts, we can imagine that each
analyst plays the role of a submodel. We seek to assign weights to
the various analysts (submodels). In the event that all pairwise
Bayes factors were consistent with one another (i.e., if
$\widehat{B}_{ij} = \widehat{B}_{il} \widehat{B}_{lj}$ for all $i,j,l
= 1, \ldots , k$), we would wish to assign relative weights according
to Bayes theorem. That is, we would wish to have
\[
\frac{b_i}{b_j} = \widehat{B}_{ij}
\]
for all $i,j = 1,\ldots ,k$. Our expression for the $b_i$ does just
this. In fact, making use of equation (\ref{eq:weight_synthesis}) and
of the consistency of the Bayes factors with one another, we have
\[
\frac{b_i}{b_j} =  \Biggl[\prod_{l=1}^k
\frac{\widehat{B}_{il}}{\widehat{B}_{jl}} \Biggr]^{1/k} =
 \Biggl[\prod_{l=1}^k \widehat{B}_{il}\widehat{B}_{lj} \Biggr]^{1/k} =
 \Biggl[\prod_{l=1}^k \widehat{B}_{ij} \Biggr]^{1/k}
= \widehat{B}_{ij}.
\]

\subsection{Model weights: Uniqueness} \label{mw_unique}
There is a sense in which our definition of
equation (\ref{eq:weight_synthesis})
is uniquely the ``correct''
means of combining information across the analysts in a broad class of
versions of the problem. We first restrict consideration to
expressions for $b_i$ which satisfy
\[
\log(b_i) = \sum_{l=1}^k  \bigl(c + d \log(\widehat{B}_{il}) \bigr)
\]
for some choice of real-valued coefficients, $c$ and $d$. This
restriction enforces linearity of the $\log(b_i)$ in the log Bayes
factors (which, in turn, are derived from log marginal likelihoods).
The restriction also ensures that common coefficients ($c$ and $d$)
are assigned, irrespective of subscripts $i$ and $l$. This is
appropriate because, in our splits, we assign the same amount of data
to each analyst, and so the same amount of data is used to compute the
log marginal likelihoods for each of the pairwise Bayes factors.
Second, to satisfy our desired property, we enforce the fixed solution
$\log(b_i / b_j) = \log (\widehat{B}_{ij} )$ when the Bayes
factors are consistent with one another. Letting $L_{ij} =
\log(\widehat{B}_{ij})$, we then have a chain of algebraic
expressions, to wit,
\begin{eqnarray*}
\log(b_i / b_j) & = & \sum_{l=1}^k  [c + d L_{il} ] -
\sum_{l=1}^k  [c + d L_{jl} ]
= d  \Biggl[\sum_{l=1}^k L_{il} - \sum_{l=1}^k L_{jl} \Biggr] \\
& = & d \sum_{l=1}^k [L_{il} + L_{lj}]
= d \sum_{l=1}^k L_{ij}.
\end{eqnarray*}
This yields the log Bayes factor comparing analyst $i$ to analyst $j$
only when $d = 1/k$, resulting in our definition of $b_i$ (up to a
multiplicative constant that drops out when deriving the relative
weights for the analysts).

\subsection{Alternative weights}
The analysts' summaries can be combined in many fashions, including
those not motivated by Bayes theorem. A simple method of this form
takes a convex combination of the analysts' summaries, but does not
update the weights. We call this method Convex Synthesis.

\section{Applications}\label{application}

In this section we describe an experiment which demonstrates the
benefits of Bayesian Synthesis and Convex Synthesis. To conduct the
experiment, we selected a data set which has been used by other
authors to illustrate the benefits of automated modeling methods. None
of us was familiar with the data set and we each received one third of
the data. This allowed us to create three pairs of analysts, with one
third of the data reserved for evaluation of the pair's synthesis.
The syntheses were compared to a variety of automated procedures. We
found that both Bayesian Synthesis and Convex Synthesis perform well.

\subsection{Ozone data}

The ozone data set consists of daily measurements of ozone
concentration and eight meteorological quantities in the Los Angeles
basin for 330 days in the year 1976. Breiman
(\citeyear{Breiman2001}) describes the
origin of the data set. The data set is contained and documented in
the software package R. The data frame contains 330 observations on
the following variables: \textit{upo3}---maximum 1-hour average upland
ozone concentration, in ppm;\footnote{Investigation of ozone standards
suggests that the units for \textit{upo}3 are actually parts per hundred
million rather than ppm. See, for example, the US EPA standards for
ground level ozone to which we return in
Section~\ref{sec:human_vs_auto}
(\href{http://www.epa.gov/ozonepollution/history.html}{http://www.epa.gov/ozonepollution/}
\href{http://www.epa.gov/ozonepollution/history.html}{history.html}
).} \textit{vdht}---Vandenberg 500 millibar height,
in meters; \textit{wdsp}---wind speed, in
miles per hour; \textit{hmdt}---humidity; \textit{sbtp}---Sandburg air base
temperature, in degrees Celsius; \textit{ibht}---inversion base height, in
feet; \textit{dgpg}---Daggett pressure gradient, in mmHg; \textit{ibtp}---inversion
base temperature, in degrees Fahrenheit; \textit{vsty}---visibility, in
miles; \textit{day}---calendar day, an integer number between 1 and
366.

Each analyst was charged with the task of constructing a Bayesian model that
can be used to predict ozone concentration. Each model should produce
a distribution for ozone concentration supported on the nonnegative
integers.

\subsubsection{The split-data analysis}

We split the data into three sets of 110 observations each, with a
complete randomization. Each of us (Analysts~1--3) received one part
of the data (data~1--3). All three analysts decided independently to
model log ozone level as a continuous variable and to produce the
agreed-upon distribution for ozone (over the positive integers) by
integrating the continuous density of the modeled variable.

\textit{Model 1.} Analyst $1$ used data set $1$ to build a model,
pursuing a strategy of first discovering which variables appeared to
be important in predicting ozone level and then determining
the forms in which the variables should enter the model.

Matrices of scatter plots of the response variable and explanatory
variables were examined. Serial dependence was investigated by
including lagged responses as explanatory variables. Several variables
(\textit{sbtp}, \textit{ibht}, \textit{vsty} and \textit{day}) appeared to be quite important,
and so were chosen to appear in the models. There was no apparent
serial dependence in the data, after adjusting for other variables.

Having identified important variables, the analyst searched for
appropriate forms. The term \textit{ibht} was modeled as four variables,
a linear term, two further variables developed to capture
nonlinearity, and an indicator for \textit{ibht}${} = 5000$, an apparent
truncation point for the variable. The indicator allows for the jump
that we expect at the truncation point and provides a~way to
incorporate additional variability at this point. The analyst used a~%
sine curve for the effect of variable \textit{day} to force it to be
periodic with period $1$~year.

After basic models
were created, the analyst reexamined variables previously judged to be
of lesser import with added variable plots and best subsets regressions.
The variable \textit{hmdt} was included as a
predictor, in a piecewise linear fashion.
The variables \textit{dgpg} (with linear and quadratic
terms) and \textit{vdht} were considered to be potential predictors.
Plots of \textit{vsty} showed a wiggly pattern of nonlinearity. Two
forms for this effect were considered---a~linear effect and a Gaussian
process centered at a linear effect. The prior on the Gaussian
process version was chosen to force the realized effect curve to be
close to linear.

Finally, eight models (all including the initial variables
and \textit{hmdt}; then the~%
$2^3$ combinations including or excluding \textit{dgpg} and \textit{vdht}
and with two forms of prior for \textit{vsty}) were selected to receive
positive probability. The prior distribution on each model was
improper, uniform for some coefficients and vague for most other
coefficients. Weights were formed for the eight models through
estimated likelihoods. Each model was updated with $99$ cases and a
predictive likelihood computed for the remaining $11$ cases. This
process was repeated $10$ times, yielding ten predictive likelihoods.
The weight given to each model was proportional to the geometric mean
of its predictive likelihoods.

\textit{Model 2.} Based on data set~2, Analyst~2 plotted log ozone
concentration and all other covariates against ``day'' to detect
evident trends. The response and the covariates were each
detrended through local fitting [by means of the \texttt{loess()}
function in R] using the variable ``day'' as a predictor. All
subsequent modeling was conducted on the residuals from these fits.

Analyst~2 believed that time proximity might constitute an important
factor and decided to specify conditional autoregressive (CAR) models
for the detrended data. Denoting the response variable by $Y$, a CAR
model takes the form $Y_t\sim \mathit{Normal}(\mu_t,\sigma^2),$ where
$\mu_t=\mathbf{X}_t^\prime\bbeta+ \bt{\theta}_t$, with $\mathbf{X}_t$
denoting a vector of covariate values at time $t$ and $\bbeta$ a
vector of model parameters. The models specified random walk priors
of order either one or two for the vector
$\bt{\theta}=(\theta_1,\ldots ,\theta_{366})^\prime$, as explained in
Thomas
et al. (\citeyear{Thomas2004}).

Analyst~2 built two models for the regression $\mathbf{X}_t^\prime\bbeta$.
The first has an intercept and four main effects
selected by means of graphical and exploratory data analysis
techniques. The second has many more predictors selected
through a stepwise procedure, starting
from the model with all main effects and two-way interactions. The
two regression models and the two CAR structures were combined to
produce four models that were averaged according to weights given in
Table~\ref{tab:wgh}.
The weights were chosen subjectively to reflect the analyst's higher
degree of confidence in simpler rather than more complicated
models. Noninformative priors were specified for the model parameters
and Winbugs was used to draw separate samples from the posterior
distributions for the four models.

%
\begin{table}
\tabcolsep=0pt
\tablewidth=265pt
\caption{Weights for Analyst 2's four component models, given data set~2.
The four component models of the mixture model
produced by Analyst~2
result from all possible combinations of
two regression models (rows)~and two CAR error structures (columns)}
\label{tab:wgh}
\begin{tabular*}{265pt}{@{\extracolsep{\fill}}lcc@{}}
\hline
&\textbf{CAR 1}&\textbf{CAR 2}\\
\hline
Main effects& 0.4 & 0.3 \\
Main effects plus interactions & 0.2 & 0.1\\
\hline
\end{tabular*}
\end{table}

\textit{Model 3.} Analyst 3 used data set 3 and applied a
modification of
Least Angle Regression [LARS; Efron
et al. (\citeyear{Efron2004})]
to fit the model: first modified
LARS was used to choose the variables to be included in the model, and
then Bayesian linear regression was implemented to quantify the
relationship between log ozone concentration and the selected
variables.

Two modifications are applied to LARS. The first is the restriction
that an interaction term can be selected only after the corresponding
main effects have entered the model. As soon as the main effects
enter, the interaction term becomes a candidate variable. The second
modification to LARS is that some variables (in this analysis,
one main effect) are forced to enter the model at the beginning of
the procedure.

Assume there are $p$ candidate main effects. Order
these variables by the strength of their correlation with the response
variable, from strongest to weakest.
Label the ordered variables $1, \ldots , p$. Suppose that variable $2$
will be forced into the model. We start with only variables $2$
through $p$ as candidate variables, and so LARS selects variable $2$.
We continue with the solution path until another variable is added.
At this point, the list of candidate variables is expanded to include
variable $1$ and the second-order term for variable $2$. A second variable
is chosen from the list of candidate variables according to the LARS
criterion. Then the second-order term for this variable and its
interaction with variable $2$ are included as candidate variables.
The above process is repeated until the solution path is completed.

Analyst $3$ used modified LARS to decide, with different forced-in
variables, the order in which variables entered the models. This
produced several sequences of models. Each sequence was examined by
$C_p$ and by differences in AIC and BIC to subjectively determine
which models were viable. A~Bayesian linear regression was computed
for each viable model, against an improper prior distribution.
Finally, BIC was used to obtain a weight for each of the four models.
With new data, both the weight for each model and the distributions of
parameters within the model were updated.

\subsubsection{Human modeling versus automated modeling}
\label{sec:human_vs_auto}

Many authors have advocated the use of automated modeling strategies,
arguing that such methods provide better predictive performance than
corresponding subjectively built models. Breiman
et al. (\citeyear{Breiman1984}) and Gu
(\citeyear{Gu2002}) analyze the ozone data with the goal of predicting log ozone
concentration. Using the methods described in their work as well as a
number of other methods, we reanalyzed the data, comparing their
predictive performance to that of the single and combined models of
Analysts 1--3.

The suite of automated methods used for comparison was chosen to span
the variety of strategies that are currently in vogue. These
strategies range from rigid strategies which select a model from a
small set of potential models and that may suffer from bias to
flexible strategies that allow an essentially arbitrary mean function
and that may overfit the data. They include both strategies that rely
on a single fit to the observed data (as in model selection) and
strategies that incorporate model averaging (whether different models
are fit to the single data set, or whether models are derived from a
collection of data sets produced from the actual data set). The
methods include both classical and Bayesian methods. Publicly
available software routines were used to implement all of the
automated methods. In general, default values were used for parameter
settings, except for the case of smoothing splines where variables
were selected using the method described in Gu
(\citeyear{Gu2002}). Specifically,
the methods investigated are those described at the end of the
\hyperref[sec1]{Introduction}.

The methods were compared on a range of goals, including those that
would naturally favor the automated analyses and those which we expect
to be difficult for the automated methods. We now step through a
brief description of the results of the comparisons.

Table~\ref{ozonet1} compares the methods in terms of prediction of log
ozone. Recall that previous analyses of these data have focused on
log ozone, and all three of the analysts also selected a log
transformation of ozone before analyzing the data. With this
transformation, a (discretized) normal likelihood appears to be
appropriate for analysis of the data. Thus, accuracy of predictions
as measured by sum of squared prediction errors provides both a
measure of the discrepancy between the predictions and the observed
data and it is directly tied to likelihood-based assessment of the
models' lack of fit.

\begin{table}
\tabcolsep=0pt
\caption{Comparison of Automatically Fitted Models with Human Models
by \textup{Sum} of Squared Errors for \textup{Log Ozone}. The row labels in
upper case indicate the modeling method under consideration. The
column labels Data set~1, 2 and~3 indicate which third of the data
was used as the test data (with the other two thirds having been
used for model building). The subcolumn labels Once and 10/10
indicate the type of prediction problem under consideration\label{ozonet1}} 
\vspace*{-5pt}
\begin{tabular*}{\textwidth}{@{\extracolsep{\fill}}lcccccc@{}}
\hline
&\multicolumn{6}{c@{}}{\textbf{Test data}}\\[-5pt]
&\multicolumn{6}{c@{}}{\hrulefill}\\
 & \multicolumn{2}{c}{\textbf{Data set 1}}& \multicolumn{2}{c}{\textbf{Data set 2}} & \multicolumn{2}{c@{}}{\textbf{Data set 3}} \\[-5pt]
& \multicolumn{2}{c}{\hrulefill}& \multicolumn{2}{c}{\hrulefill} & \multicolumn{2}{c@{}}{\hrulefill}\\
\textbf{Updating method} & \multicolumn{1}{c}{\textbf{Once}} & \multicolumn{1}{c}{\textbf{10/10}}
& \multicolumn{1}{c}{\textbf{Once}} & \multicolumn{1}{c}{\textbf{10/10}} &
\multicolumn{1}{c}{\textbf{Once}} & \multicolumn{1}{c@{}}{\textbf{10/10}}\\
\hline
ANALYST 1 & -- & -- & 12.31 &
12.43 & 14.65 & 14.03 \\
ANALYST 2 & 17.96 & 17.59 & -- &
-- & 15.66 & 15.78 \\
ANALYST 3 & 15.96 & 16.07 & 14.21 & 14.32 & -- &
-- \\
[4pt]
MN. HMN. PR. ERR. & 16.96 &
   16.83 & 13.26 &
   13.38 & 15.15 &
   14.91 \\
BAYES SYNTH. & 15.98 &
16.29 & 11.93
& 11.98 &
13.39 & 14.32
\\
CONVEX SYNTH. & 15.98 &
15.81 & 11.93
& 12.08 &
13.39 & 13.27
\\
[4pt]
CART & 27.51 & 21.29 & 17.87 & 18.72 & 19.37 & 17.81 \\
BAYES TREE & 28.56 & 25.39 & 22.12 & 19.76 & 20.04 & 21.39 \\
BAGGED CART & 19.66 & 19.02 & 14.91 & 14.22 & 16.32 & 15.64 \\
BART & 13.10&12.31&11.40&11.06&13.21&12.87\\
SS & 19.75 & 20.15 & 17.21 & 17.23 & 17.63 & 15.27 \\
LARS & 21.33 & 21.55 & 17.36 & 19.17 & 19.40 & 28.50 \\
LASSO & 21.37 & 21.76 & 16.76 & 19.12 & 20.50 & 28.64 \\
FWD STGW & 21.12 & 21.11 & 17.20 & 19.44 &20.50 & 28.28 \\
BMA & 21.96 & 21.89 & 17.61 & 17.67 & 16.90 & 16.31 \\
AIC & 20.84 & 19.88 & 16.91 & 16.29 & 16.75 & 15.76 \\
BIC & 21.51 & 20.78 & 17.47 & 16.90 & 16.75 & 15.76 \\
\hline
\end{tabular*}
\tabnotetext[]{t1}{Note that, to improve readability, this
table summarizes sum of squared errors, while Table~\protect\ref{table:ozone_original_scale}
summarizes mean squared errors.}
\vspace*{-3pt}
\end{table}

The table contains six comparisons. For each comparison, one split of
the data is reserved as test data, with the other two splits used to
fit the models. In addition, two versions of the prediction problem
were investigated. The first is a static prediction problem, the
latter a sequential prediction problem. For the static problem, the
training data were used to develop the model. A~prediction was made
for each case in the test data, and the measure of fit was computed.
We refer to this as making a prediction ``once and for all.'' For the
sequential problem, we randomly partitioned the test data into 11 sets
of 10 cases each. The model was fit to the training data, and a~%
prediction made for the first set of cases in the test data. The
model was updated (getting the posterior distributions both within and
across models) based on the first set of cases in the test data, and
predictions made for the second set of cases. This procedure was
continued, updating the model on successively larger sets of data and
making predictions for the next set of cases, until the test data were
exhausted. We used the same partition of the test data (in the same
order) to evaluate each of the methods. We refer to this as ``ten by
ten'' evaluation.

Table~\ref{ozonet1} contains rows for the ``Mean Human Prediction
Error,'' for ``Baye\-sian Synthesis'' and for ``Convex Synthesis.''
The Mean Human Prediction Error is defined by selecting an analyst
at random to make predictions. The measure of fit is the mean of the
two analysts' measures. Bayesian Synthesis implements the method
of Section~\ref{framework}, combining the two analysts eligible to
make predictions for the test data. The initial weights given to each
analyst are equal to $1/2$. When updating ten by ten, the weights
adjust, based on the relative performance of the analysts' models.
The predictions were taken to be the posterior predictive means.
Convex Synthesis uses the same procedure as Bayesian
Synthesis, but maintains a constant weight of $1/2$ for each analyst.
Because the initial weights are equal to $1/2$ for both Bayesian and
Convex Syntheses, the once and for all updating yields the same
results in both cases. For the 10 by 10 updating the final weights
under Bayesian Synthesis are 0.019 for Analyst~2 and 0.981 for
Analyst~3 when predicting data set~1, 1.000 for Analyst~1 and 0.000
for Analyst~3 when predicting data set~2, and 0.990 for Analyst~1 and
0.010 for Analyst~2 when predicting data set~3.\looseness=-1

Table~\ref{ozonet1} shows the success of data splitting and of human
modeling. We first note that the Mean Human Prediction Error
provides a better predictive fit than do any of the classical
automated methods. Mean human prediction error corresponds to
randomly selecting an analyst to develop a model. This comparison
establishes the benefit of subjective modeling.

Second, we turn to the main purpose of the experiment---to see whether
Bayesian Synthesis outperforms rival methods. In every instance
(excepting BART), we find that the method does outperform competing
procedures. Bayesian Synthesis and Convex Synthesis yield much
smaller predictive mean square errors than do any of the automated
methods. The predictive mean square error is also smaller than the
Mean Human Prediction Error. Bayesian Synthesis and Convex
Synthesis outperform both human analysts in five of the six
comparisons and is virtually as accurate as the better analyst in the
remaining one. The comparisons also show the magnitude of the benefit
to human modeling. The differences between the bulk of the automated
techniques are considerably smaller than the differences between these
automated techniques and the syntheses. As noted above, Convex
Synthesis and Bayesian Synthesis are identical for ``Once''; Convex
Synthesis performs better than Bayesian Synthesis for two out of the
three 10 by 10 updatings.

Third, the comparison between the static and sequential problems
shows, on the whole, a modest benefit to continually updating the
model. It also makes clear the dominant role that modeling plays in
effective prediction---building a better model (more precisely, a
better collection of models) is far more important than having a bit
more data with which to update the model.

Table~\ref{table:ozone_original_scale} repeats the comparisons in
Table~\ref{ozonet1}, but with ozone replacing log ozone as the
response. Providing predictions for the human analysts, the Mean
Human Prediction Error, Bayesian Synthesis,
Convex Synthesis and BART is straightforward, because
for these methods an MCMC Bayesian summary of the posterior
distribution is available. In these cases, the models developed for
log ozone imply corresponding models for ozone: The prediction for a
case is given by its predictive mean. In terms of mean squared error
of prediction, BART does best, followed by Convex Synthesis,
followed by Bayesian Synthesis, which in
turn outperforms all human analysts.

To provide predictions for the automated methods (other than BART), we
faced a choice between use of the method with strongly skewed
likelihood or ad-hoc correction of a model developed on the log ozone
scale. The latter route generally provided better performance, and
Table~\ref{table:ozone_original_scale} presents these results. To
provide predictions, a model was developed for log ozone, the
prediction, say, $\widehat{y}$, was obtained for each case, as was an
in-sample estimate of mean squared error, say, $\widehat{\mathit{MSE}}$. The
prediction for ozone was taken to be $\exp\{\widehat{y} + 0.5
\widehat{\mathit{MSE}}\}$. The results in
Table~\ref{table:ozone_original_scale} are in general accord with
those of Table~\ref{ozonet1}. The main difference is that the
superiority of Bayesian Synthesis and Convex Synthesis
relative to other methods has
decreased.

\begin{table}
\tabcolsep=0pt
\caption{Comparison of Automatically Fitted Models with Human Models
by \textup{Mean} Squared Errors for \textup{Ozone}. The row labels in upper
case indicate the modeling method under consideration. The column
labels Data set~1, 2 and~3 indicate which third of the data was
used as the test data (with the other two thirds having been used
for model building). The subcolumn labels Once and 10/10 indicate
the type of prediction problem under consideration\label{table:ozone_original_scale}} 
\vspace*{-5pt}
\begin{tabular*}{\textwidth}{@{\extracolsep{\fill}}lcccccc@{}}
\hline
&\multicolumn{6}{c@{}}{\textbf{Test data}}\\[-5pt]
&\multicolumn{6}{c@{}}{\hrulefill}\\
 & \multicolumn{2}{c}{\textbf{Data set 1}}& \multicolumn{2}{c}{\textbf{Data set 2}} & \multicolumn{2}{c@{}}{\textbf{Data set 3}} \\[-5pt]
& \multicolumn{2}{c}{\hrulefill}& \multicolumn{2}{c}{\hrulefill} & \multicolumn{2}{c@{}}{\hrulefill}\\
\textbf{Updating method} & \multicolumn{1}{c}{\textbf{Once}} & \multicolumn{1}{c}{\textbf{10/10}}
& \multicolumn{1}{c}{\textbf{Once}} & \multicolumn{1}{c}{\textbf{10/10}} &
\multicolumn{1}{c}{\textbf{Once}} & \multicolumn{1}{c@{}}{\textbf{10/10}}\\
\hline
ANALYST 1 & -- & -- &
15.05 & 15.00 & 25.00 & 25.60\\
ANALYST 2 &
16.48 & 16.00
& -- &
-- &
22.37 & 22.94\\
ANALYST 3 &
14.90 & 15.29 & 15.92 & 16.08
&
-- & -- \\
[4pt]
MN. HMN. PR. ERR.&
16.08 &   15.68 & 15.52 &   15.54 & 23.72 &   23.81\\
BAYES SYNTH. &
14.52 & 14.83 & 13.10 & 14.38 & 21.44 & 22.36\\
CONVEX SYNTH. &
14.52 & 14.29 & 13.10 & 13.30 & 21.44 & 21.62
\\
[4pt]
CART & 25.50 & 20.43 & 25.20 & 20.88 & 24.21 & 22.18\\
BAYES TREE & 24.90 & 21.72 & 29.81 & 24.80 & 22.94 & 24.70\\
BAGGED CART& 18.58 & 17.81 & 18.92 & 17.14 & 18.75 & 18.84\\
BART & 12.26 & 11.87 & 13.09 & 12.34 & 18.07 & 18.60\\
SS & 18.32 & 18.32 & 26.42 & 15.92 & 23.72 & 21.44\\
LARS & 17.89 & 19.62 & 15.13 & 17.98 & 25.20 & 28.09\\
LASSO & 18.66 & 19.01 & 15.76 & 18.40 & 27.35 & 28.20\\
FWD STGW & 18.32 & 18.66 & 14.98 & 17.98 & 27.35 & 27.77\\
BMA & 18.40 & 20.88 & 15.13 & 16.16 & 21.62 & 22.18\\
AIC & 17.64 & 17.89 & 15.13 & 15.13 & 20.52 & 20.52\\
BIC & 18.15 & 18.40 & 15.44 & 15.44 & 20.52 & 20.52\\
\hline
\end{tabular*}
\tabnotetext[]{t2}{Note that, to improve readability, this
table summarizes mean squared errors, while Table~%
\protect\ref{table:ozone_original_scale} summarizes sum of squared errors.}
\vspace*{-5pt}
\end{table}

Table~\ref{table:ozone_classification_error} examines forecasts of
ozone threshold exceedance. State and federal regulations provide
limitations on ozone. There are a number of ways in which ozone
thresholds can be violated, including a high peak ozone concentration
during a day and an excessive mean ozone concentration over an
extended period of time. These thresholds have varied over time, and
there has been a general downward trend in the standards. We focus on
the maximum 1-hour average standard of 0.08~ppm which was in effect
from 1971--1979. We have taken this to be 8 in units of \textit{upo}3. With
each method, a forecast (exceed or not) is made for each day in the
test data set. The table presents the number of incorrect forecasts.

For human models, combinations of human models and BART, creating the
forecast is straightforward. The model provides a predictive
distribution for ozone concentration. If the predictive probability
of exceedance is greater than 0.5, the forecast is ``exceed''; if less
than 0.5, the forecast is ``not exceed.'' The automated methods are
more difficult to deal with. For these methods, we faced a choice of
attempting to directly model ozone exceedance or to model some other
quantity and then extract a forecast of ozone exceedance. The latter
proved to be a more effective strategy. The forecasts for these
methods are based on whether the point prediction for log ozone
exceeds the threshold of $\log(8.5)$. If the point prediction exceeds
$\log(8.5)$, the forecast is for exceed; if not, the forecast is ``not
exceed.'' Convex Synthesis does the best on this task, edging
Bayesian Synthesis in the 10 by 10 updating, with BART and the Mean
Human Prediction Error close behind. The other methods lag
substantially.

\begin{table}
\tabcolsep=0pt
\caption{Classification errors (false positive plus false
negative) for forecasts of ozone threshold exceedance, with
threshold equal to 8 units of \textit{upo}3. The row labels in upper case
indicate the modeling method under consideration. The column labels
Data set~1, 2 and~3 indicate which third of the data was used as
the test data (with the other two thirds having been used for model
building). The column label Total refers to the total
classification errors over the three test data sets. The subcolumn
labels Once and 10/10 indicate the type of prediction problem
under consideration. The observed numbers~of exceedances for Data
sets~1, 2 and~3 were 58, 60 and 67, respectively}\label{table:ozone_classification_error}
\begin{tabular*}{\textwidth}{@{\extracolsep{\fill}}ld{2.1}d{2.0}d{2.1}d{2.1}d{2.0}d{2.0}d{2.0}d{2.1}@{}}
\hline
&\multicolumn{8}{c@{}}{\textbf{Test data}}\\[-5pt]
&\multicolumn{8}{c@{}}{\hrulefill}\\
 & \multicolumn{2}{c}{\textbf{Data set 1}}& \multicolumn{2}{c}{\textbf{Data set 2}}
 & \multicolumn{2}{c}{\textbf{Data set 3}}&\multicolumn{2}{c@{}}{\textbf{Total}} \\[-5pt]
& \multicolumn{2}{c}{\hrulefill}& \multicolumn{2}{c}{\hrulefill}&\multicolumn{2}{c}{\hrulefill} & \multicolumn{2}{c@{}}{\hrulefill}\\
\textbf{Updating method} & \multicolumn{1}{c}{\textbf{Once}} & \multicolumn{1}{c}{\textbf{10/10}}
& \multicolumn{1}{c}{\textbf{Once}} & \multicolumn{1}{c}{\textbf{10/10}}& \multicolumn{1}{c}{\textbf{Once}} & \multicolumn{1}{c}{\textbf{10/10}} &
\multicolumn{1}{c}{\textbf{Once}} & \multicolumn{1}{c@{}}{\textbf{10/10}}\\
\hline
ANALYST 1 & \multicolumn{1}{c}{--} & \multicolumn{1}{c}{--}
& 14& 14& 18& 17
& \multicolumn{1}{c}{--} & \multicolumn{1}{c@{}}{--}
\\
ANALYST 2 & 11 & 10 & \multicolumn{1}{c}{--} & \multicolumn{1}{c}{--}
& 14 & 15 & \multicolumn{1}{c}{--} & \multicolumn{1}{c@{}}{--}
\\
ANALYST 3 & 10 & 10 & 11 & 11 &
\multicolumn{1}{c}{--} & \multicolumn{1}{c}{--}
& \multicolumn{1}{c}{--} & \multicolumn{1}{c@{}}{--}
\\ [4pt]
MN. HMN. PR. ERR. & 10.5     & 10 &
12.5     & 12.5     &
16 & 16
& 39&38.5
\\
BAYES SYNTH.
& 10 & 10 &
12 & 13 &
15 & 14 & 37 & 37
\\
CONVEX SYNTH.
& 10 & 8 &
12 & 11 &
15 & 15 & 37&34
\\
[4pt]
CART &
21 & 13 & 15 & 16 & 17 & 17 & 53 & 46\\
BAYES TREE &
23 & 17 & 16 & 15 & 21 & 20 & 60 & 52\\
BAGGED CART &
17 & 14 & 11 & 12 & 19 & 17 & 47 & 43\\
BART &
11 & 11 & 11 & 12 & 17 & 15 & 39 & 38\\
SS &
16 & 15 & 14 & 11 & 17 & 18 & 47 & 44\\
LARS &
17 & 17 & 12 & 18 & 16 & 19 & 45 & 54\\
LASSO &
17 & 17 & 12 & 18 & 16 & 19 & 45 & 54\\
FWD STGW &
16 & 17 & 12 & 17 & 16 & 22 & 44 & 56\\
BMA &
16 & 17 & 15 & 13 & 16 & 17 & 47 & 47\\
AIC &
15 & 15 & 13 & 12 & 16 & 16 & 44 & 43\\
BIC &
15 & 15 & 15 & 14 & 16 & 16 & 46 & 45\\
\hline
\end{tabular*}
\end{table}

In addition to the comparisons presented here, we have examined
several other potential evaluations. Some of these appear in
Yu
(\citeyear{Yu2006}). Overall, we find a substantial advantage for the human
models and for BART.

\subsubsection{One Bayesian versus Bayesian Synthesis}

The previous comparative exercises demonstrate that the syntheses
provide an improvement over the individual Bayesian. In nearly all
instances,
Bayesian Synthesis and Convex Synthesis have performed better than the
Mean Human Prediction Error. This alone leads us to recommend routine
use of our techniques. In this section we examine two more
comparative exercises, both of which show the syntheses to be
preferable to individual analysts and to the Mean Human Prediction
Error. The comparisons are ``once and for all'' comparisons and so
Bayesian Synthesis and Convex Synthesis have identical performance.
We also include BART in this comparison because it is a Bayesian
method and so leads to noncontroversial predictive variances and
predictive intervals.

The focus of these additional comparisons is calibration of the
posterior predictive distribution. To look at this issue, we make two
comparisons. The first is accuracy of coverage rates of prediction
intervals. We form 90\% prediction intervals for the three data sets
as before. The intervals are central predictive probability
intervals, cutting off 5\% of the predictive distribution in each
tail. Table~\ref{table:ozone_calibration} presents these results
under \% cvg. We find generally good agreement with nominal coverage
levels, with the syntheses and BART performing a little better
than individual analysts.

\begin{table}
\tabcolsep=0pt
\caption{Calibration of the
posterior predictive distribution for log ozone. The row labels in
upper case indicate the modeling method under consideration. The
column labels Data set~1, 2 and~3 indicate which third of the data
was used as the test data. In the top table, the subcolumns labeled
Var contain the estimated variances of the predictive
distribution, averaged over the 110 predicted cases in the data set
and the subcolumns labeled MSE contain the mean squared errors
of prediction for the 110 predicted cases. The subcolumns labeled
\% cvg contain the observed coverage rates of 90\% prediction
intervals. In the bottom table, the results in the top table are
averaged over the three data~sets. The column labeled optimism
contains the ratio of the average MSE to the average~variance.~A~value of the ratio exceeding one corresponds to an overly optimistic~assessment~of~predictive~accuracy}\label{table:ozone_calibration}
\begin{tabular*}{\textwidth}{@{\extracolsep{\fill}}lccccccccc@{}}
\hline
&\multicolumn{9}{c@{}}{\textbf{Test data}}\\[-5pt]
&\multicolumn{9}{c@{}}{\hrulefill}\\
 & \multicolumn{3}{c}{\textbf{Data set 1}}& \multicolumn{3}{c}{\textbf{Data set 2}} & \multicolumn{3}{c@{}}{\textbf{Data set 3}} \\[-5pt]
& \multicolumn{3}{c}{\hrulefill}& \multicolumn{3}{c}{\hrulefill} & \multicolumn{3}{c@{}}{\hrulefill}\\
  & \multicolumn{1}{c}{\textbf{Var}} & \multicolumn{1}{c}{\textbf{MSE}}& \multicolumn{1}{c}{\textbf{\% cvg}}
& \multicolumn{1}{c}{\textbf{Var}} & \multicolumn{1}{c}{\textbf{MSE}}& \multicolumn{1}{c}{\textbf{\% cvg}}
& \multicolumn{1}{c}{\textbf{Var}} & \multicolumn{1}{c}{\textbf{MSE}}& \multicolumn{1}{c}{\textbf{\% cvg}}\\
\hline
ANALYST 1 & -- & -- &
-- &
0.119 & 0.112 & 92.73 & 0.109 & 0.133 & 83.64
\\
ANALYST 2 &
0.135 & 0.163 & 85.45 &
-- & -- &
-- &
0.142 & 0.142 & 90.00
\\
ANALYST 3 &
0.115 & 0.145 & 85.45 & 0.123 & 0.129 & 88.18 &
-- & -- &
--
\\
[4pt]
MN. HMN. PR. ERR. & 0.125 & 0.154 & 85.45 & 0.121 & 0.121 & 90.45 &
0.126 & 0.138 & 86.82
\\
SYNTHESES & 0.134 & 0.145 & 86.36 & 0.133 & 0.109 & 92.73 & 0.140
& 0.140 & 90.00
\\
[4pt]
BART & 0.112 & 0.119 & 90.00 & 0.115 & 0.104 & 92.73 &
0.102 & 0.120 & 86.36\\
\hline
\end{tabular*}
\begin{tabular*}{\textwidth}{@{\extracolsep{\fill}}lcccc@{}}
\hline
&\multicolumn{3}{c}{\textbf{Average}} & \\[-5pt]
&\multicolumn{3}{c}{\hrulefill} &\multicolumn{1}{c@{}}{\textbf{Optimism}}\\
&
\multicolumn{1}{c}{\textbf{Var}} & \multicolumn{1}{c}{\textbf{MSE}} &
\multicolumn{1}{c}{\textbf{\% cvg}}& \multicolumn{1}{c@{}}{\textbf{MSE/Var}}
\\
\hline
ANALYST 1 &
0.114 & 0.123 & 88.18 & 1.077
\\
ANALYST 2 &
0.139 & 0.153 & 87.73 & 1.102
\\
ANALYST 3 &
0.119 & 0.137 & 86.82 & 1.153
\\
[4pt]
MN. HMN. PR. ERR. &
0.124 & 0.138 & 87.58 & 1.111
\\
SYNTHESES &
0.135 & 0.131 & 89.70 & 0.968
\\
[4pt]
BART &
0.110 & 0.114 & 89.70 & 1.040\\
\hline
\end{tabular*}
\end{table}

\begin{figure}

\includegraphics{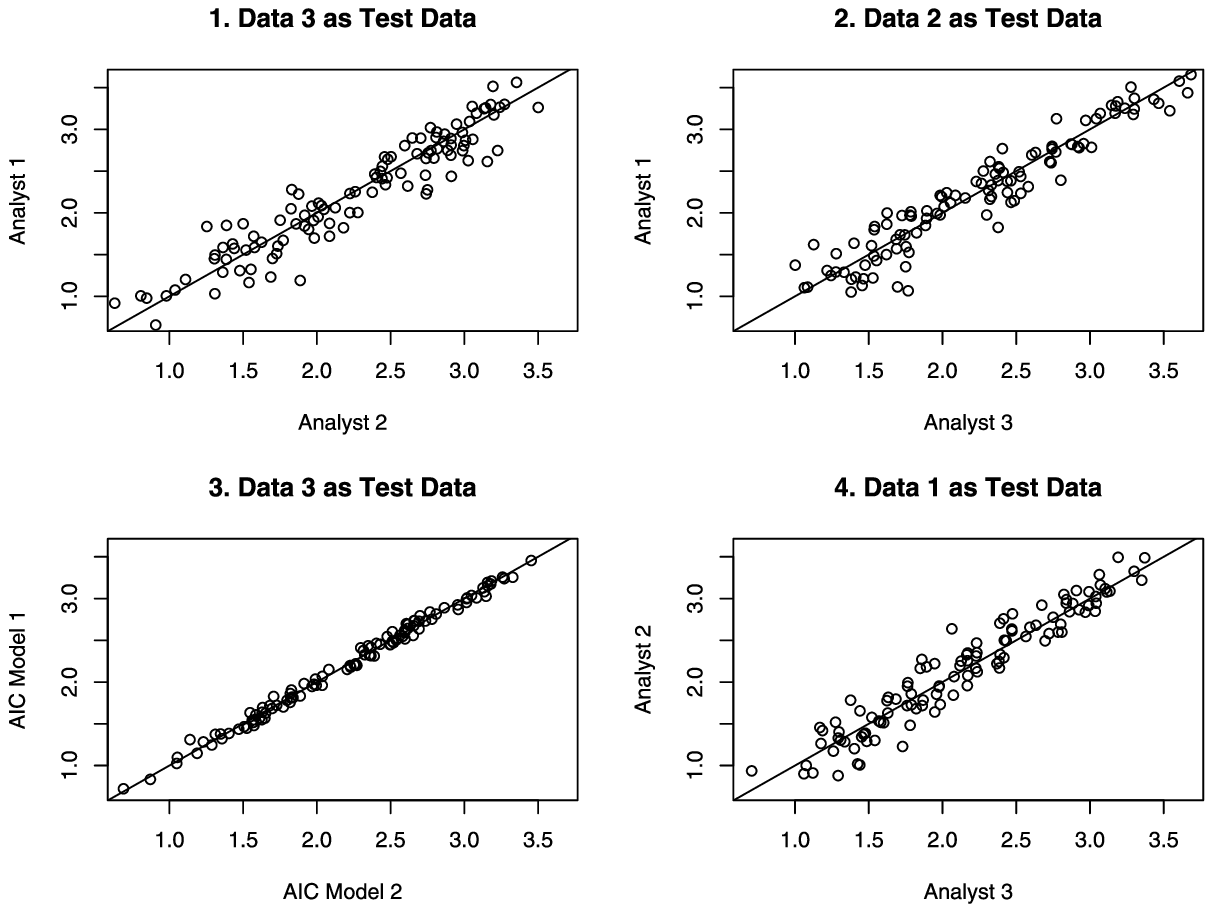}

\caption{Plots showing the differences in out-of-sample model
predictions for the models developed by the human analysts and the
models developed by AIC using various splits of the
data.}\label{ozone_diff}
\end{figure}

The second comparison is of internal and external measures of
accuracy. For these measures, we focus on the predictive distribution
for log ozone. Under a Bayesian model, the expected squared departure
from the predictive mean is the predictive variance. Thus, as an
internal measure of accuracy, we use the variance of the predictive
distribution, averaged over the 110 predicted cases. As an external
measure of accuracy, we use the mean squared error of prediction. The
results are presented in Table~\ref{table:ozone_calibration}. We note
that the analysts' internal estimates systematically understate the
actual variation, while the syntheses and BART produce nearly
equivalent internal and external measures of accuracy. The ratio of
MSE to Var is a measure of the optimism of the Bayesian. When this
ratio exceeds 1, the Bayesian is overly optimistic. We computed these
ratios based on the average MSE and variance over the three data sets.
The ratios, summarized in Table~\ref{table:ozone_calibration}, exceed
one for all methods other than the syntheses.

\subsubsection{Why the syntheses work}
We next turn to an explanation of the benefits of model synthesis.
The syntheses, indeed all Bayesian model averaging, provide the
greatest benefits when the models to be synthesized provide different
predictions. It is here that averaging allows one to make a~different
prediction than either model, and it is here that further information
collected in data allows the posterior weights given to different
models to select the better model. The benefits of bagging/averaging
models arising from relatively stable procedures such as AIC, BIC and
SS are minimal (results not presented in the tables), because the bulk
of the bagged models provide the same or similar predictions.
Figure~\ref{ozone_diff} shows that differences in predictions from
different analysts show more variation than do differences from
different AIC models.

The results outlined in Table~\ref{ozonet1} show clearly that there
are large benefits stemming from human modeling with additional
improvements attributable to the syntheses. Interestingly, large
benefits can also ensue from synthesis of a human and an automatically
fitted model, as evidenced by the summaries presented in Yu
(\citeyear{Yu2006}).
This is in part due to the fact that the predictions produced by human
and automatically fitted models are typically different. Also, the
gains appear to be more sizable when the human models are synthesized
with methods based on the creation of new variables (e.g., Smoothing
Spline, CART, Bagged CART, BART) than when they are synthesized with
methods based on regressions with the original variables (e.g., AIC,
BIC, BMA, LARS, LASSO, Forward Stagewise). Overall, the empirical
results indicate that
the predictions produced by the syntheses usually outperform
the predictions of the single constituent elements and inherit many of
the performance properties of the best of the constituent elements.

Across our set of comparisons, Convex Synthesis has outperformed
Baye\-sian Synthesis by a modest margin. We find this surprising, as
our expectation was that Bayesian Synthesis, by updating the weights,
would tilt the predictions toward the analyst with the better fitting
model, resulting in better predictive performance. We do not have a
definitive explanation for this behavior, but we do conjecture that it
is due in part to shortcomings of all of the analysts' models. The
``data-generating mechanism'' is, presumably, not captured by any of
the analysts. As the analysts' models are not nested within one
another, a convex combination of the analysts' predictions enlarges
the space of predictions. It is plausible that this expanded space
includes models that fit better than those of any individual analyst,
producing the observed results. A related discussion, where the truth
is presumed to lie within the convex hull of a collection of models,
appears in Kim
and Kim (\citeyear{Kim2004}).

\section{Discussion and further research}\label{discussion}

In this paper we propose Bayesian Synthesis and Convex Synthesis, a
new paradigm for Bayesian data analysis. The paradigm is motivated by
the concern that using a set of data both to develop a model and to
subsequently fit the model with the same data violates the spirit of
Bayes theorem. The paradigm has been developed with an eye to which
parts of a modeling effort appear to be stable---model development by a
single analyst---and which appear to yield highly variable
results---model development by different analysts. Tapping into the
variable parts of an analysis while retaining enough information to
preserve stability of the other parts of the analysis allows us to
obtain the greatest benefits of Bayesian model averaging. This also
provides us with a more appropriate accounting of model uncertainty.

We have explored the new paradigm experimentally.
Yu
(\citeyear{Yu2006}) contains a~theoretical motivation for the work,
providing an ensemble of theorems that justifies split-data analyses.
Experimentally, the ozone data analysis shows the remarkable
benefits that accrue to subjective modeling and the further benefits
that follow from synthesizing subjective models across analysts.

In practice, it is more costly and time-consuming to produce several
subjective analyses than a single one, so when should this method be
employed? We recommend use of this method when the amount of
available data is sufficient to produce split data sets that are
informative, and when the problem is important enough to justify the
involvement of several analysts. Examples of such situations include
efficacy and safety studies in large clinical trials, post-enumeration
adjustment of the census, industrial research and development, and
large marketing surveys. Situations for which the method is not
recommended are those where real-time predictions are needed, as is
the case for internet searches, target recognition and on-line
quality control, unless the components of the synthesis can be built
ahead of time. In the latter case, the type of synthesis to be
employed will need to avoid the expense of a formal Bayesian updating
of the weights.

This work raises several issues. One issue is how to most effectively
split the data. In this work, we have focused on partitioning the
data set with randomization playing a dominant role. An alternative
route is to allow overlapping splits of the data, so that each analyst
receives a more than $1/k$ fraction of the data. We expect
overlapping splits to be of most use when data sets are small or when
they contain large numbers of potential predictors. Overlapping
splits also allow us to benefit from the modeling efforts of a larger
set of analysts. The theoretical results in Yu
(\citeyear{Yu2006}) address these
overlapping splits.

A second issue is the development of prototypical problems so that a~%
precise methodology can be specified depending on the goal(s) of the
analysis and the type of data collected. Investigation of these
problems will give us more guidance on how to split the data and on
what restrictions to place on the Bayesian summaries.

A third issue is application of the methodology with non-Bayesian
components. The benefits of averaging nonstable or different models
applies more broadly than in the Bayesian setting. Noting differences
between the models built by CART and by the information criteria, one
could average them as well. However, without a Bayesian summary and
with incomplete likelihoods, model synthesis becomes somewhat more
ad-hoc. Convex Synthesis provides one such simple method which could
be implemented with fixed weights, as we have done here, or with
weights determined by some predefined rule. Natural routes to pursue
include the prequential approach [e.g., Dawid
and Vovk (\citeyear{Dawid1999})] and
predictive model selection [e.g., Laud
and Ibrahim (\citeyear{Laud1995})].

\section*{Acknowledgments} The authors would like to thank the Editor
and an Associate Editor for
insightful comments that improved the paper.


%

\printaddresses

\end{document}